\documentclass[preprint,12pt]{elsarticle}

\usepackage{graphicx}
\usepackage{amssymb}
\usepackage{amsthm}

\usepackage{graphicx} 
\usepackage{dcolumn} 
\usepackage{bm} 
\usepackage{hyperref} 
\usepackage{amsmath}
\usepackage[capitalise]{cleveref} 
\crefname{figure}{Figure}{Figures} 
\Crefname{figure}{Figure}{Figures} 
\crefname{table}{Table}{Tables}
\Crefname{table}{Table}{Tables}
\crefname{equation}{Eq.}{Eq.}
\Crefname{equation}{Equation}{Equations}
\crefname{appendix}{}{}
\Crefname{appendix}{}{}
\usepackage{siunitx} 
\usepackage[caption=false]{subfig} 
\usepackage{verbatim} 

\usepackage[textsize=tiny]{todonotes}

\usepackage{lineno}

\biboptions{comma,round, sort&compress}

\journal{Computational Materials Science}

\begin{document}
\begin{frontmatter}

\title{Phase behaviour of (Ti:Mo)S$_2$ binary alloys arising from electron-lattice coupling}

\author[1]{Andrea Silva\corref{cor1}}
\ead{a.silva@soton.ac.uk}

\author[1,3]{Tomas Polcar}%
\ead{t.polcar@soton.ac.uk}

\author[1,4]{Denis Kramer}%
\ead{d.kramer@soton.ac.uk}

\cortext[cor1]{Corresponding author}
\address[1]{Engineering Materials, University of Southampton}
\address[3]{Advanced Materials Group, Department of Control Engineering, Faculty of Electrical Engineering, Czech Technical University in Prague (CTU)}
\address[4]{Mechanical Engineering, Helmut Schmidt University, Hamburg, Germany}

\date{\today} 

\begin{abstract}
While 2D materials attract considerable interests for their exotic electronic and mechanical properties, their phase behaviour is still largely not understood.
%
%
This work focuses on (Mo:Ti)S$_2$ binary alloys which have captured the interest of the tribology community for their good performance in solid lubrication applications and whose chemistry and crystallography is still debated.
Using electronic structures calculations and statistical mechanics we predict a phase-separating behaviour for the system and trace its origin to the energetics of the $d$-band manifold due to crystal field splitting.
Our predicted solubility limits as a function of temperature are in accordance with experimental data and demonstrate the utility of this protocol in understanding and designing TMD alloys.
\end{abstract}


\end{frontmatter}


\section{\label{sec:intro}Introduction}
The unique structural and electronic properties of 2D materials attract considerable interest.
Applications range from nanostructured electronics such as FET channels~\cite{Wang2012, Chhowalla2013} to Li- and Na-ions batteries~\cite{Ma2015, Larson2018} and solid lubrication~\cite{Roberts1990,Polcar2011a,Vazirisereshk2019}.
While theoretical and experimental studies tend to focus on the electronic and mechanical properties, the phase behaviour of low dimensional materials, equally important to inform synthesis strategies and understand service life, remains an open question~\cite{Ektarawong2019}.

This work focuses on the phase behaviour of a compound that has captured the interest of the tribology community: (Mo:Ti)S$_2$ alloys have been identified as a promising material with enhanced tribological properties both by experiments and computational investigations. 
Mesoscale experiments report that Ti-doped composite coatings show better resistance to oxidation compared to pristine MoS$_2$, while preserving low friction coefficients~\cite{Fox1999}. 
Recent computational work~\cite{Cammarata2015b,Cammarata} has put forward an argument to rationalise the good frictional behaviour in terms of the vibrational properties: the low-frequency optical phonon modes taken to be associated with the perfect shear of two layers are extrapolated along the sliding path and taken as an indication of low energy barriers for \textit{sliding modes}. 
By studying this descriptor across the transition-metal dichalcogenide (TMD) chemical space, the authors identified layered 2H-Ti$_{1/4}$Mo$_{3/4}$S$_2$, where a quarter of TM sites within the TM-S$_2$ layers is occupied by Ti, as a candidate material with enhanced frictional properties compared with other analysed TMDs.
Albeit the interest attracted by this compound, the exact structure and chemistry are still debated.
To experimentally realise this computationally engineered chemistry, it would be advantageous for Ti$_{1/4}$Mo$_{3/4}$S$_2$ to be thermodynamically stable within the Ti-Mo-S chemical space.
However, the reported mechanically stability in Density Functional Theory (DFT) calculations is by itself insufficient to assess the thermodynamic viability.
Experiments do not provide a definite answer either. On the one hand, studies on thin films synthesised via magneto-sputtering suggest Ti is not fully integrated within the TM-S$_2$ planes~\cite{Fox1999,Renevier2000}.
On the grounds of the measured ratio between chemical composition and film hardness, the authors conclude that if Ti were fully incorporated in the layered structure, the resulting properties of the material, including hardness, should deviate more from pristine MoS$_2$ than is observed.
On the other hand, a separate study of Ti-doped MoS$_2$ nanostructures concluded that Ti is incorporated within the TM-S$_2$ planes~\cite{Hsu2001}.
The authors disregarded possible intercalation between the layers due to the absence of distortion along the interlayer $c$ axis, suggesting Ti is embedded in the planes.

In an attempt to elucidate the thermodynamic viability of layered (Mo:Ti)S$_2$ compounds including 2H-Ti$_{1/4}$Mo$_{3/4}$S$_2$, we computationally investigate the energetics and thermodynamics of site substitutions to identify ordered phases and solubility limits along the full (Mo:Ti)S$_2$ pseudo-binary line within the Mo-Ti-S phase space.
\begin{figure}[ht]
\centering
  \includegraphics[width=0.85\textwidth]{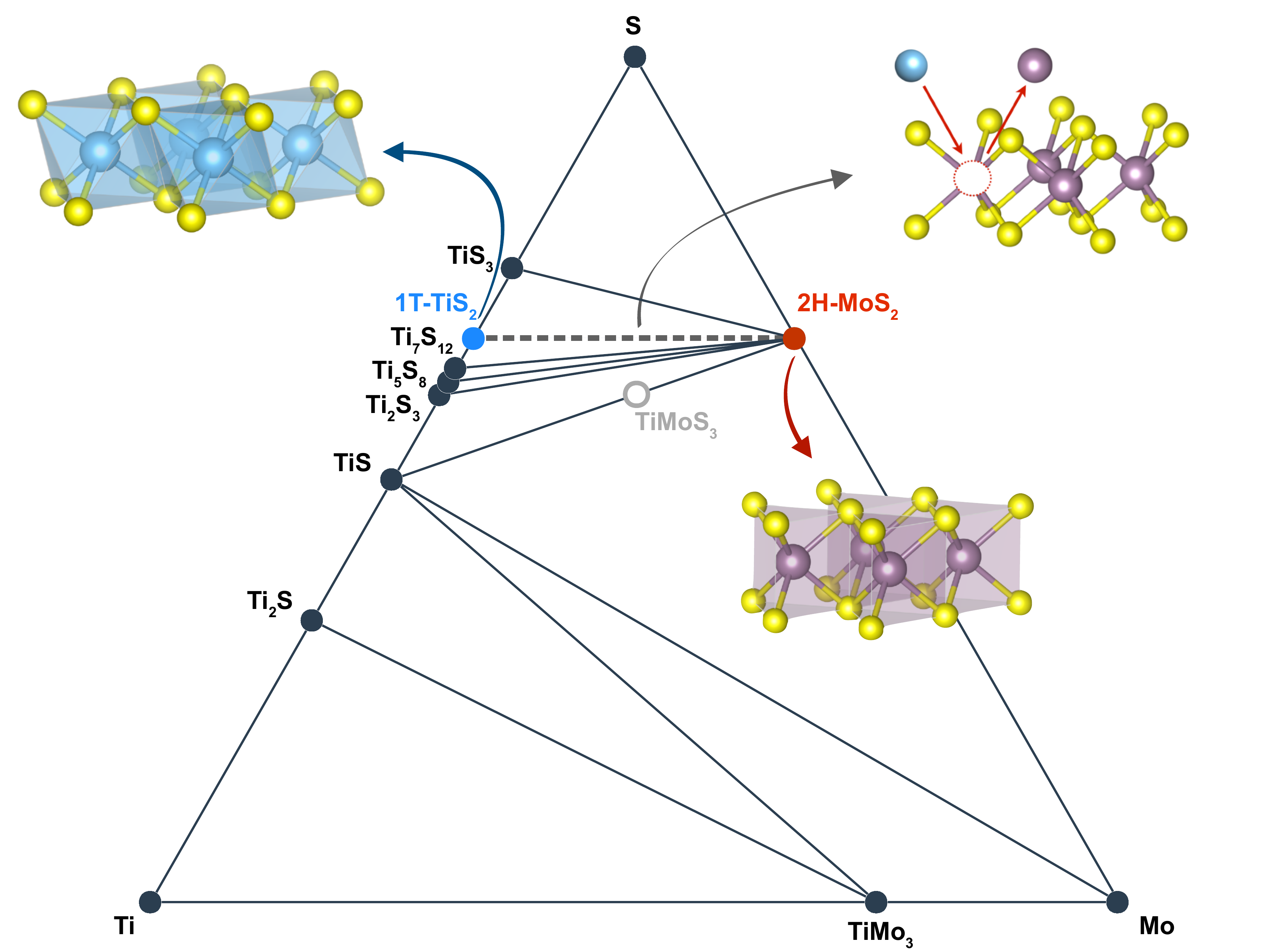} 
  \caption{\label{fig:alloy_sketch}
  Ternary phase diagram of stoichiometric compounds in the Mo-Ti-S space based on formation energies provided by the Materials Project database.
  Insets on the left and right show the coordination of the pristine compounds, 1T-TiS$_2$ and 2H-MoS$_2$, respectively.
  The binary alloy system studied here is represented by the dashed line and the substitutional process sketched at the bottom.
}
\end{figure}
\Cref{fig:alloy_sketch} visualises the considered part of the ternary phase space. 
We consider only compounds Ti$_x$Mo$_{1-x}$S$_2$ within the two host structures suggested by the binary sulfides.
The ternary plot in \cref{fig:alloy_sketch} reports stable configurations and tie lines in the chemical space according to the Materials Project Database~\cite{Ong2008,Jain2013a}.
The only ternary compound reported is TiMoS$_3$, an unstable non-layered structure with formation energy $\SI{0.04}{eV}$ above the convex hull.
To the best of our knowledge, there are no stable ternaries reported in the phase field MoS$_2$-TiS$_2$-Ti$_7$S$_12$ relevant here, and no attempt was made to computationally search for unknown ternaries in this region of the chemical space.

The two end-members are:
\begin{enumerate}
    \item 1T-TiS$_2$ is a semi-metal that crystallises in a layered compound with space group $P\bar{3}m1$.
    The structure is depicted on the left side in \cref{fig:alloy_sketch}.
    Triangular Ti layers are sandwiched by triangular chalcogenide planes, mutually rotated by $\ang{60}$, resulting in octahedrally coordinated Ti. 
    The 1T prefix indicates that TiS$_2$  planes are stacked in an AA fashion~\cite{Kolobov2016}.
    \item 2H-MoS$_2$ is a semiconductor with an indirect bandgap of $\SI{1.3}{eV}$ and crystallises in a layered structure with space group $P6_3/mmc$. 
    Mo planes are sandwiched by S planes that are not rotated relative to each other, resulting in prismatic coordination of Mo, as sketched on the right side in \cref{fig:alloy_sketch}.
    The 2H prefix designates an AB stacking order of MoS$_2$ layers~\cite{Kolobov2016}.
\end{enumerate}
To explore the effects of system dimensionality on the phase stability, the systems are also studied in purely monolayer (ML) form.
No staking order is present in the ML case and the pristine compounds are referred to as 1T-TiS$_2$ and 1H-MoS$_2$.
%

\section{\label{sec:methods}Methods}
\subsection{Cluster Expansion method}\label{sec:CE}
The combinatorial problem of cation ordering within a crystal structure is addressed using the Cluster Expansion (CE) formalism~\cite{Connolly1983}. 
The CE approach, effectively, coarse grains all degrees of freedom of the system until only the site occupancy on a lattice remains.
%
%
The energy of a configuration $\sigma$ of the lattice is expanded in terms of  clusters functions $\Phi_\alpha(\sigma)$ representing different interaction patterns
\begin{equation}\label{eq:CE}
    E(\sigma)= \sum_\alpha J_\alpha \Phi_\alpha(\sigma),
\end{equation}
where the sum runs over the distinct orbits $\alpha$.
An orbit $\alpha$ is a set of symmetry-equivalent clusters $\{\beta\}$. 
The cluster functions $\Phi_\alpha = \left< \prod_{i\in \beta} \sigma_i \right>$ are averages over all clusters $\beta$ within an orbit $\alpha$.
The strength of each interaction is given by the constants $J_\alpha$, termed Effective Cluster Interaction (ECI).
Even though \cref{eq:CE} is exact in the limit of an infinite number of orbits $\alpha$, the series must be truncated for any practical application.
In most system, the shortsighted nature of interactions allows a satisfactory description of the behaviour retaining only a few clusters~\cite{Wolverton1987,Sanchez2017}.
%
%
The utility of a CE is twofold.
Once a CE-model has been built, this provides a fast-evaluating Hamiltonian describing the configurational space of the system that allows to efficiently search for new ground-state orderings.
Iterating this search and refining the CE model until the underlying interaction is described correctly, one has obtained an Ising-like lattice Hamiltonian that enables the evaluation of thermodynamic properties via Monte Carlo simulations.

The Alloy Theory Automated Toolkit (ATAT)~\cite{VandeWalle2002} has been used to construct cluster expansions. 
The series \cref{eq:CE} is truncated at figures of four vertices and cross-validation is used to select the most predictive model over a training set. 
The expansion is considered converged once the ground states predicted by the CE agree with DFT calculations and the error on predicted energies is deemed negligible.

\subsection{First-principles calculations}
Total energy calculations of geometries along the Ti$_x$Mo$_{1-x}$S$_2$ tie-line are performed using DFT within the Projector Augmented-Wave (PAW) framework~\cite{Blochl1994} as implemented in the Vienna \textit{Ab initio} Simulation Package (VASP)~\cite{Kresse1993,Kresse1999}. 
Exchange-correlation effects are modelled using the Strongly Constrained Appropriately Normed (SCAN) functional~\cite{Sun2015}.
The subtle van der Waals interactions coupling the layers in bulk systems are described using the non-local kernel correction rVV10~\cite{Vydrov2010}.
This combination has proven to accurately describe layered materials and complex geometries~\cite{Peng2016}.

A plane wave cut-off of $\SI{800}{eV}$ was adopted for all DFT calculations; the Brillouin zone of the pristine compounds in their primitive cell was sampled using an $11\times11\times11$ mesh and the number of $k$ points per reciprocal atom was kept constant for larger supercells.
For the ML geometries, layers of TMD are separated by $\SI{20}{\text{\AA}}$, ensuring there is no interaction between periodic images.
In order to obtain accurate total energies, cell vectors and atomic positions were relaxed with a convergence cut-off of $\SI{0.5}{meV/atom}$. 

\subsection{Monte Carlo simulations}
To investigate the effect of temperature and configurational entropy on the system, Monte Carlo simulations were performed using the CE Hamiltonians and the EMC routine of the ATAT package~\cite{VandeWalle2002b}.
The simulations are carried out in the semi grand-canonical ensemble, where the chemical potential $\mu$, number of lattice sites $N$ and temperature $T$ are fixed while concentration $x$ and energy $E$ can fluctuate. 

The supercells used in the calculation are reported in \cref{tab:CE_data}.
All ground states of the system are stabilized within the chosen chemical-potential range, spanned in steps of $\Delta\mu = \SI{0.01}{eV}$.
Temperature is varied as function of its inverse $\beta=1/T$ between $\beta_1=1/\SI{100}{K}$ and $\beta_1=1/\SI{8000}{K}$ in steps $\Delta\beta=\SI{1e-4}{}$.
This ensures a high sampling density at low temperatures while conveniently enabling the ideal solid solution case as a high-temperature staring point.
%
%
Each MC simulation is considered converged once concentration fluctuations are less than the threshold of $\Delta x = 5 \cdot 10^{-3}$.

\begin{table}[h]
\centering
  \begin{tabular}{l|cccc}
  Host & Training set size & Clusters $\Phi_\alpha$& CV [eV] & MC cell size \\
  \hline
  2H bulk & 57 & 19 & 0.009 & 37x37x8 \\
  1T bulk & 113 & 31 & 0.052 & 21x21x11 \\
  1H ML & 39 & 8 & 0.016 & - \\
  1T ML & 46 & 37 & 0.083 & -
  \end{tabular}
  \caption{\label{tab:CE_data}
  Training set and convergence of the CE in the trigonal prismatic 2H and octahedral 1T hosts, for bulk and ML geometries.}
\end{table}

\section{Results }\label{sec:results}
\subsection{Crystallography and Cluster Expansion}
The lattice parameters of the pristine compounds 2H-MoS$_2$ and 1T-TiS$_2$ as obtained from calculations are reported in \cref{tab:lattice_param} and compared with experimental crystallographic data. 
The DFT-SCAN with rVV10 correction describes the in-plane bonding accurately and stacking lattice constants are in good agreement with experiment, which indicates that the rVV10 kernel accurately captures the cohesive inter-layer interactions.
\begin{table}
\centering
    \begin{tabular}{l|cccc}
      Compound & $a  [\text{\AA}]$ & $c [\text{\AA}]$ & Method & Reference \\
      \hline
      2H-MoS$_2$ & 3.168  &  12.5 & DFT-SCAN & This work  \\ 
                 & 3.161  &  12.3 & experimental     &~\cite{Dickinson1923}  \\
      1T-TiS$_2$ & 3.409  & 5.75   & DFT-SCAN & This work \\
                 & 3.410  & 5.70   & experimental     &~\cite{Chianelli1975}  \\
    \end{tabular}
  \caption{\label{tab:lattice_param}
  Intralayer $a$ and interlayer $c$ lattice parameters from simulations and experiments.
  Bulk lattice parameter $a$ is within 0.03\% and 0.2\% of the experimental measured value, while the interlayer one $c$ is within 0.7\% and 1.8\% for TiS$_2$ and MoS$_2$, respectively.}
\end{table}

\begin{figure}[h]
    \centering
	\includegraphics[width=0.85\textwidth]{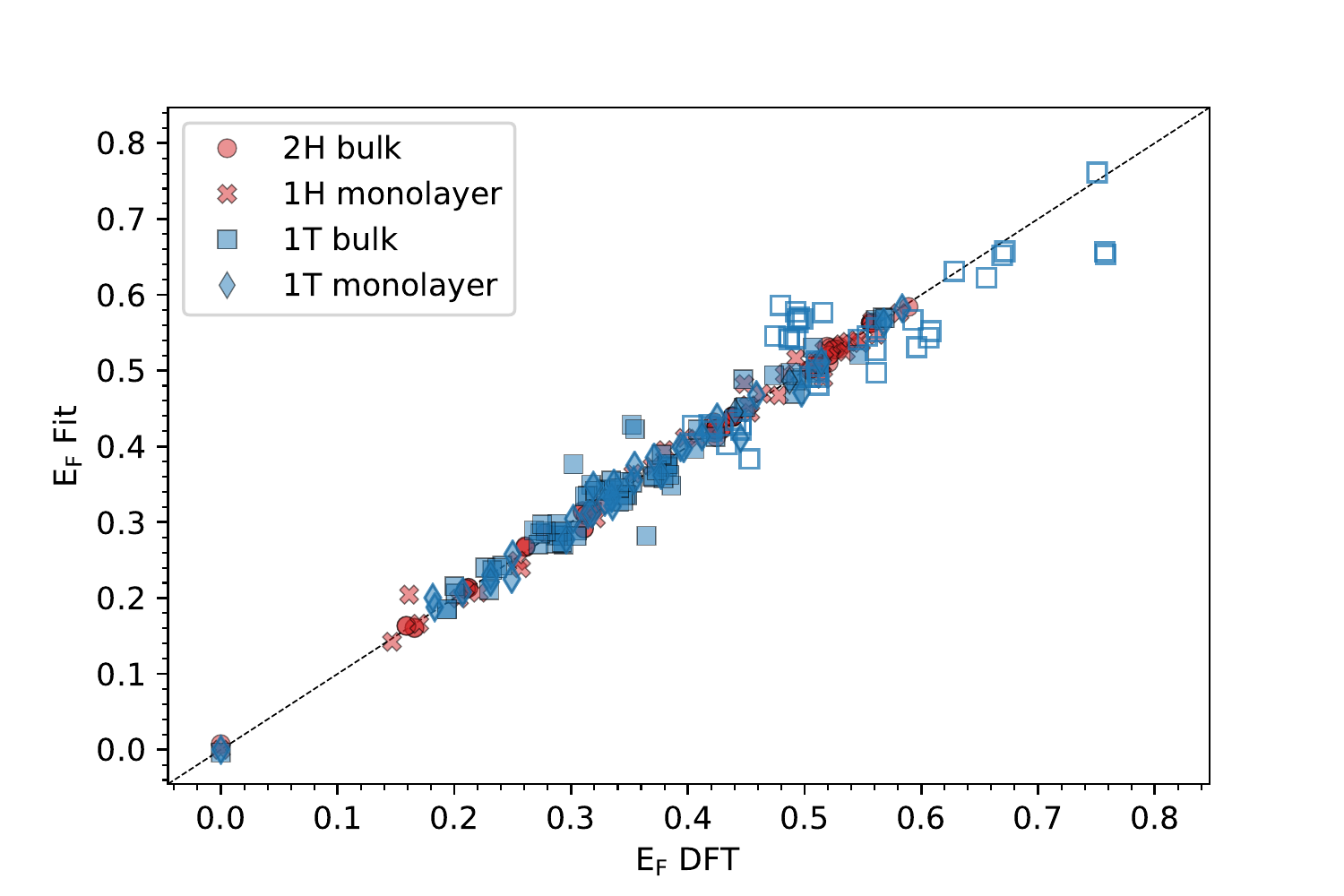} 
	\caption{\label{fig:fit_res}
	Fitted versus DFT-computed formation energies in all systems.
	In a perfect fit, all points would lie on the bisector, shown as a dashed black line.
	Empty squares refer to configuration $x>0.6$, which have a different weight in the fit as explained in \cref{sec:phase_stab}.
	}
\end{figure}

Separate CE Hamiltonians were built for the trigonal-prismatic (H) and the octahedral (T) hosts. 
Two datasets of total energy calculations, one per host, were used to train the ECIs, which are reported in greater detail in 
\cref{sec:ECI}.
A dataset comprising around 50 structures was sufficient to bring CE and DFT into agreement for the trigonal-prismatic host, while a little over a hundred configurations were needed for the octahedral host as reported in \cref{tab:CE_data}.
The CE for the trigonal-prismatic host was build using the full concentration range $x \in [0, 1]$, while the CE model for the octahedral host was biased to accurately reproduce the ground-states only in $x \in [0, 0.6]$ as explained in \cref{sec:phase_stab}.
As reported in \cref{fig:fit_res}, energies predicted by CE models agree with the DFT-computed ones: the average error is $0.8 \pm 6.9~\SI{}{meV/site}$ for the 2H host and $-2 \pm 24~\SI{}{meV/site}$ for 1T host.
Considering that the energy landscape is spanned by the formation energy of end-members in the non-native host, which is of the order of 0.5 eV, this error is negligible in the description of the energetics defining most phase diagrams.
The error in the 1T host is comparable to the thermal energy at room temperature $k_\mathrm{B}T_\mathrm{room}=\SI{25}{meV}$, making it relevant for low-temperature simulations at low concentration $x$, near the 1T-TiS$_2$ end-member.
However, in the high-temperature portion of the phase diagram, which is the one of interest here, this error becomes negligible.

\subsection{Convex Hull}\label{sec:phase_stab}
\begin{figure}[h]
\centering
  \includegraphics[width=0.85\textwidth]{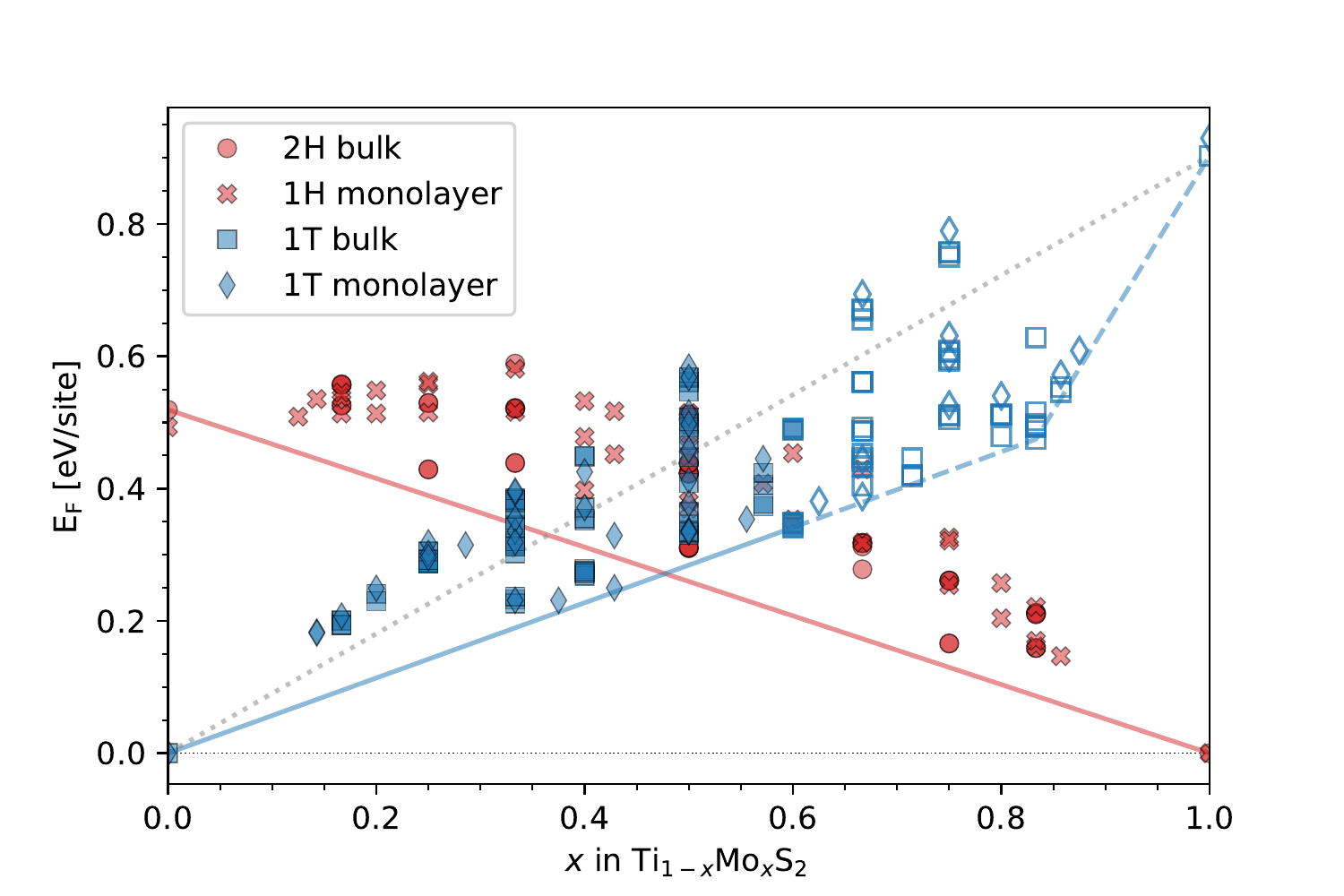}
  \caption{ \label{fig:hull}
  The figure reports DFT-computed energies for the H host in bulk (red circles) and ML form (red crosses) and T host in bulk (blue squares) and ML (blue diamonds).
  Hollow symbols mark the 1T configuration at $x>0.6$, where the CE model is not required to reproduce the right ground state.
  The dashed grey line connects the end-members formation energies of the T host.
  Red and blue solid lines show the convex hull within the H and T hosts, respectively.
  The black dotted line marks the zero-formation energy limit.
  }
\end{figure}

In order to understand stability across different hosts, the formation energy of a configuration is defined as follows
\begin{align}
    \label{eq:formation_en}
    E_\mathrm{F}^{\mathrm{host}}(x) &= 
        E^{\mathrm{host}}(\mathrm{Mo}_x\mathrm{Ti}_{1-x}\mathrm{S}_2) \nonumber \\
        &- x E^{\mathrm{2H}}(\mathrm{MoS_{2}}) 
        - (1-x) E^{\mathrm{1T}}(\mathrm{TiS_{2}}),
\end{align}
where $E^{\mathrm{host}}(\mathrm{Ti}_{1-x}\mathrm{Mo}_x\mathrm{S}_2)$ is the energy per TM site of the configuration $\sigma(x)$ at concentration $x$, and remaining terms are the total energy of the pristine compound in the ground-state hosts.
%
%
%
Formation energies from \cref{eq:formation_en} are reported in \cref{fig:hull} for bulk and ML (ML) in both octahedral 1T and trigonal prismatic 1H and 2H hosts.
The line connecting the end-member in each host (solid red for H host and dashed grey for T host in \cref{fig:hull}) represents the energy of ideal solid-solution with negligible interactions between the fraction $x$ of sites occupied by Mo and the remaining Ti sites.
Points lying below this line represent stable configurations in the given host while points over it mark energetically unfavourable regions, where Mo-rich and Ti-rich parts are segregated within the same host geometry.
Finally, stable structures across both hosts would show negative formation energies, lying below the black dotted line in \cref{fig:hull}, but no such configuration has been found.

The prismatic host is not receptive to alloying. 
All training set configurations lie above the line connecting the end-members (red symbols in \cref{fig:hull}), indicating a high energy penalty for Ti in prismatic coordination.
%
%
Since no ordered arrangement of the two species yields an energy gain, Mo and Ti ions within the 2H host segregate at $\SI{0}{K}$.
Conversely, energy favourable orderings are found within the octahedral host.
Several training set configurations lie below the ideal solid-solution line, as shown by blue marks below the grey dotted line in \cref{fig:hull}.
In particular the CE iterative search identifies ground-state orderings at $x=0.60$ (Mo$_{3/5}$Ti$_{2/5}$S$_2$) and at $x=0.83$ (Mo$_{5/6}$Ti$_{1/6}$S$_2$).
According to the convex hull in \cref{fig:hull}, the octahedral T host is favourable until $x\approx 0.5$, after which the prismatic H host becomes more stable.
%
%
This concentration-depend stability between the two hosts at $\SI{0}{K}$ and the large distortions occurring in the T host for $x > 0.6$, as reported in 
\cref{sec:host_dist}, 
motivated the decision to constrain the training set of the 1T model: CE Hamiltonians are required to reproduce correctly the 1T ground-state only within the range $x \in [0, 0.6]$.
This is because at higher concentrations the system will prefer the H host and the CE has difficulties to capture the energetics of large lattice distortions~\cite{VandeWalle2002}.

Since no part of either convex hull lies below the zero-formation energy line of the composite-host system (black dotted line in \cref{fig:hull}), the system is deemed phase-separating at $\SI{0}{K}$: the lowest-energy configuration at any concentration comprises two separate regions of 1T-TiS$_2$ and 2H-MoS$_2$.
Only at finite temperature, entropic effects could stabilise the presence of mixed-concentration configurations within a single host.

\subsection{Miscibility at higher T}
The CE Hamiltonians for the 2H and 1T bulk system trained with the data-points in \cref{fig:hull} was used to run finite-temperature MC simulations to build free-energy curves for each host that govern the stability of the system once temperature and configurational entropy are introduced.
Since the MC simulations are carried out in a semi grand-canonical ensemble, only single-phase regions of the phase diagram are directly explored by the simulations and two-phase equilibrium regions are inferred.
Then the multi-host free-energy surface is obtained by a double tangent construction: at a fixed temperature, the free energy surface of each host is built with the two-tangent construction and the convex hull of the resulting two surfaces yields the total free energy $F(x, T)$.
%
%

\begin{figure}[h]
\centering
	\includegraphics[width=0.85\textwidth]{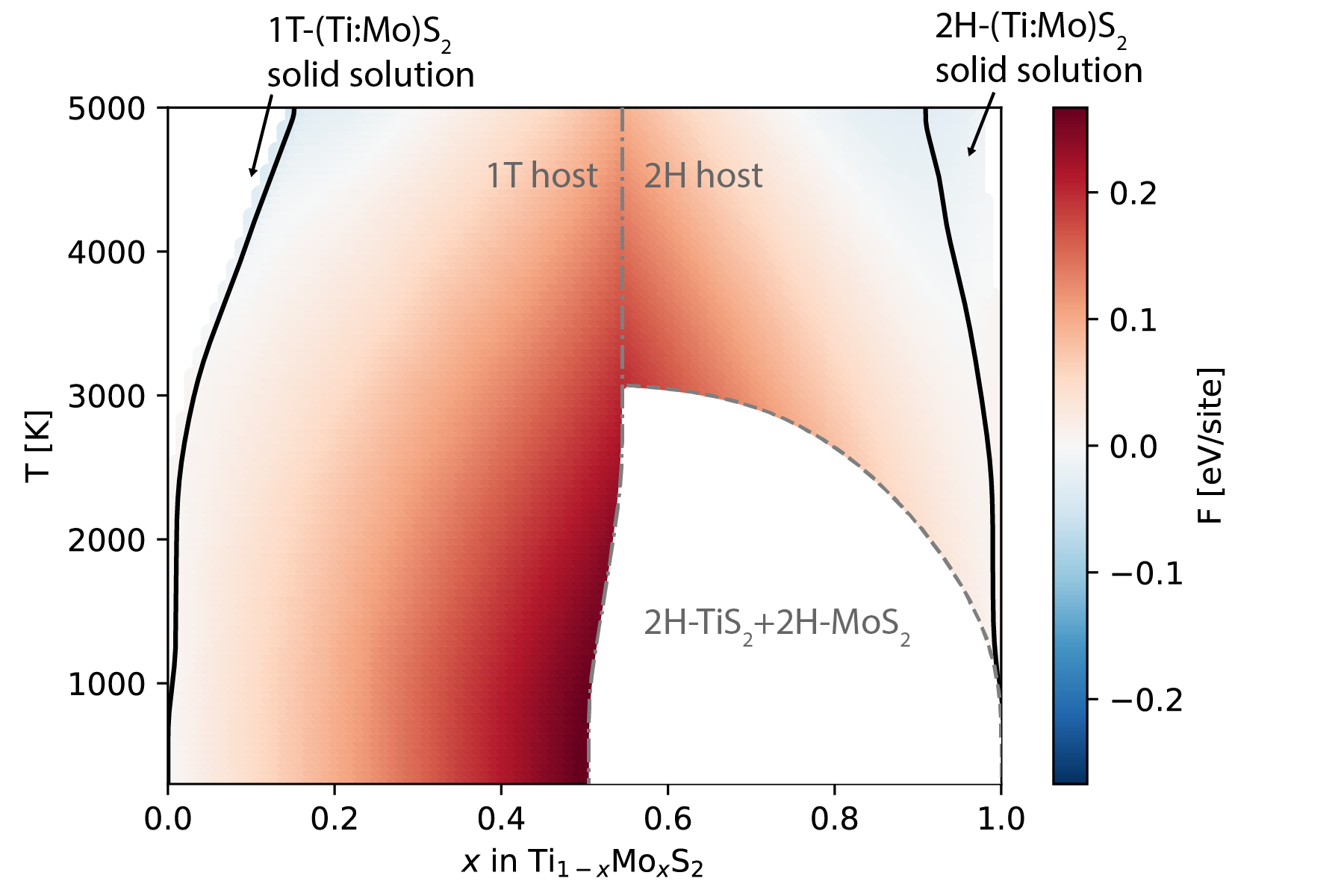}
	\caption{\label{fig:phase_diag}
	Phase diagram of the combined-host system showing temperature $T$ versus equilibrium concentration $x$ in MC simulations.
	Colours report the value of the free energy $F(x, T)$, linearly interpolated between the MC points.
	The solid black line highlights the phase boundary between solid solution in T and H hosts from phase separation into 2H-MoS$_2$ and 1T-TiS$_2$.
	%
	Grey lines indicate host predominance (lowest energy host) within the two-phase region with the dash-dotted grey line separating the 1T-host and 2H-host predominance, as indicated by the grey labels.
	The white region between dash-dotted and dashed grey lines indicates phase separation within the 2H host not accessible by semi-grand canonical MC simulations.
	}
\end{figure}

The phase diagram of the system across 2H and 1T hosts is thus divided in phase separating and solid-solution regions.
%
The color scheme in \cref{fig:phase_diag} shows the value of the multi-host Helmholtz free-energy $F(x, T)$ as function of concentration $x$ and temperature $T$.
Considering the thermodynamic behaviour of each host independently (i.e., excluding phase separation into different hosts), the T host shows solid-solution behaviour already at room temperature, while phase-separation of Ti and Mo is observed within the prismatic 2H host up to $T=\SI{3000}{K}$ at $x=0.5$. Similar to the zero temperature data shown in \cref{fig:hull}, the lower energy host changes from 1T for lower concentrations of Ti to 2H at about $x\simeq 0.5$ with a slight shift to higher $x$ at elevated temperatures.

Finally, the solid black line in \cref{fig:phase_diag} indicates the phase boundaries between two-phase separation into 1T and 2H hosts (central region) and solid solution in a single host (left-most and right-most regions) and shows that in a realistic temperature range the system is completely phase separating, with configurational entropy stabilizing only small-percentage doping around the two end-members.

\section{Discussion}
\subsection{Stabilisation mechanism in octahedral host}\label{sec:CF_stab}
\begin{figure}[h]
\centering
	\subfloat[]{\includegraphics[width=0.45\textwidth]{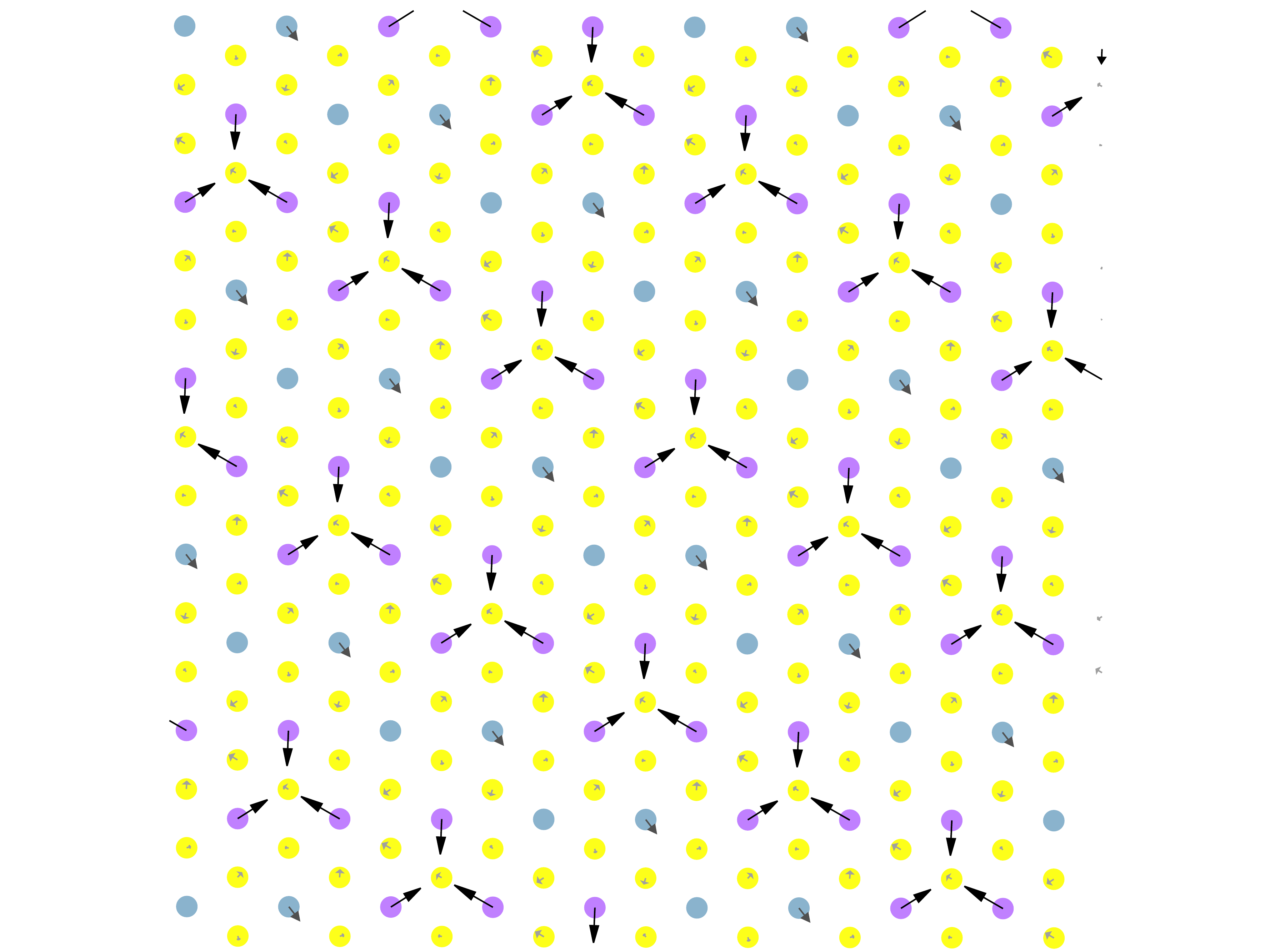}\label{fig:GS_dist}}
	\quad 
	\subfloat[]{\includegraphics[width=0.45\textwidth]{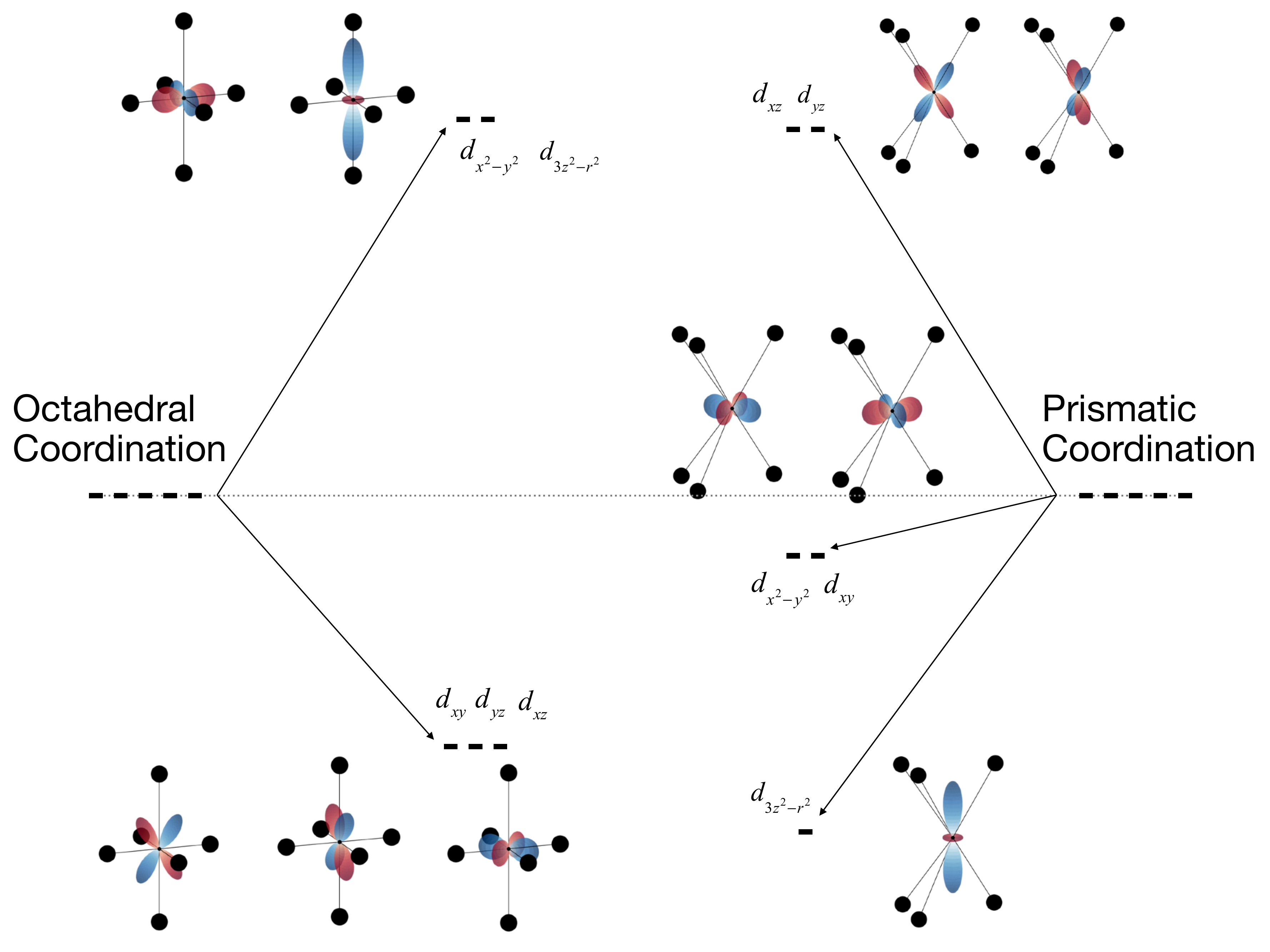}\label{fig:CFS_sketch}}
	\caption{
	\label{fig:distGS_CFS}
    (a) Azimuthal view of the starting, perfect octahedral Mo$_{3/5}$Ti$_{2/5}$S$_2$ ordering at $x=0.6$ .
    Purple, blue and yellow circles represent Mo, Ti and S ions, respectively.
    The distortion stabilising the geometry is shown with arrows, whose length and shade are proportional to the magnitude of the displacement.
	(b) Sketch of the octahedral (left) and prismatic (right) energy levels in the CF splitting picture.
	Insets beside energy levels depict the corresponding hydrogen-like orbitals on the transition metals site, surrounded by sulfur ions in the respective coordination.
	%
    }
\end{figure}

Even though it does not lead to overall-stable geometries, it is interesting to understand the origin of the stabilization mechanism leading to the ordered configuration Mo$_{3/5}$Ti$_{2/5}$S$_2$ at $x=0.6$ in the 1T host.
The ordering is sketched in \Cref{fig:GS_dist}.
The configuration consists of a striped arrangement composed of three staggered rows of Mo and two rows of Ti.
As shown by the displacement arrows in \cref{fig:GS_dist} Ti cations retain their position, locally preserving octahedral coordination, while Mo clusters distort the host, locally breaking the symmetry.
This distortion around Mo ions can be understood qualitatively with  Crystal Field (CF) theory and Kramer's theorem~\cite{Huheey1993}.
The CF description has an intuitive physical interpretation but only leads to a qualitative description of the present system, because the transition metal-chalcogenide bond shows a degree of covalency while the CF model assumes purely ionic bonding.
%
%

The CF resulting from octahedral coordination splits the five degenerate  $d$ orbitals of the isolated TM into two energy manifolds, as depicted on the left-hand side of \cref{fig:CFS_sketch}.
The lower energy $t_{2g}$ manifold consists of three orbitals, while the higher energy $e_g$ manifold comprises two orbitals. 
Prismatic coordination, on the other hand, leads to three manifolds with only the $d_{z^2}$ orbital contributing to the lowest energy manifold. 

Assigning Ti a formal valence of 4+, neither crystal field will result in an energetic advantage for the $d^0$ ion, because the triply degenerate $t_{2g}$ states of the octahedral environment and the $d_{z^2}$ states of the prismatic environment are both empty.
Hence, octahedral coordination is favoured as it provides the most efficient packing~\cite{Kertesz1984}.
%
%
On the other hand, Mo$^{4+}$ is a $d^2$ ion leading to the partial occupation of the $t_{2g}$ manifolds. 
%
%
According to Kramer's theorem, the system will, therefore, tend to lower its symmetry through Jahn-Teller (JT) distortions to break the degeneracy of $t_{2g}$ and lowers the total energy.
%
%
%
The lattice is thus divided into non-JT-active sites, i.e. Ti rows, and JT-active sites, composed of the Mo triplets clustering together.
The same mechanism cannot occur in the MoS$_2$ native prismatic coordination, as the low-energy CF level is non-degenerate for the $d^0$ configuration of Ti$^{4+}$ as well as the $d^2$ configuration of Mo$^{4+}$ ions and, thus, Kramer's theorem does not apply.

The slightly higher solubility of Mo in TiS$_2$ compared to Ti in MoS$_2$ seen in \cref{fig:phase_diag} is, therefore, attributed to the additional stability obtained from JT activity of Mo$^{4+}$ within the octahedral environment of the 1H host.

\subsection{Generalised principle for 2D TMDs design}
The dimensionality and vdW interactions do not affect the phase stability of the alloy system.
As \cref{fig:hull} reports, in both hosts, the convex hull of the bulk system and their ML counterpart present the same character.
This finding has important implications for future studies: since the phase behaviour is unchanged from the bulk case, one could extrapolate from ML phase diagrams to the bulk.
Extrapolating from monolayer formation energies to bulk without information on interlayer interactions yields errors in the prediction of the order of stacking faults energies.
Stacking fault energies are typically around $\SI{10}{meV}$ for TMDs~\cite{Irving2017a,Levita2014a} and other 2D sheets bounded by vdW dispersion~\cite{Losi2020,Reguzzoni2012} and $\SI{50}{meV}$ for transition metal oxides~\cite{Kaufman2019}.
This argument can be extended to most TMD families and layered materials in general, provided that vdW forces are the only inter-layer interactions, i.e. no Coulomb or magnetic interaction are present~\cite{Ramasubramaniam2013,Calandra2018}.

Moreover, the electron-lattice stabilisation mechanism presented in \cref{sec:CF_stab} could occur in other TMD-based compounds and must be taken into account when designing similar alloys.
The JT-based distortion lowers the energy of the ground-state configurations at $x=0.60$ in \cref{fig:hull} by about $\SI{100}{meV}$ compared to the ideal-solid solution limit.
Even though this energy gain is not enough to redefine the multi-host convex hull in the (Mo:Ti)S$_2$ system, it could lead to ground-state orderings in other similar alloys, if the formation energy of the end-members in both hosts is low enough or zero, when end-members share the same ground-state host~\cite{Worsdale2015}.

\subsection{Comparison with experimental data}
The phase diagram in \cref{fig:phase_diag} contains useful information from a synthesis point of view, allowing estimation of the maximum doping fraction at a given temperature.
For example, at $T=\SI{1200}{K}$ the maximum fraction of substituted Ti should be around $1\%$. 
Considering that the melting point of pristine MoS$_2$ and Mo-Ti metallic alloys are reported to be around $\SI{1700}{K}$ and $\SI{2000}{K}$~\cite{Baker1998}, respectively, it should be in principles possible to observe such doped configurations experimentally by high-temperature synthesis routes and subsequent quenching to inhibit the phase segregation at a lower temperature.
This prediction is consistent with the results of Hsu~\textit{et al.}~\cite{Hsu2001}, where energy dispersive X-ray analysis detected the presence of a small amount of Ti in 2H-MoS$_2$-based nanostructures obtained by mixing Mo-Ti powder and H$_2$S at $\SI{1200}{K}$, while the inter-layer lattice constant measured from the High-Resolution Transmission Electron Microscopy and X-ray diffraction fail to show an expansion, which would be indicative of Ti ions intercalated between MoS$_2$ sheets.

\section{Conclusion}
We presented the (Ti:Mo)S$_2$ phase diagram that results from considering TM substitutions within the native hosts of the pristine compounds.
Our model based on electronic-structure calculations and statistical mechanics predicts full phase separation in the system across hosts and the solubility limits inferred from our MC simulations are in agreement with the high-temperature synthesis of Ti-doped 2H-MoS$_2$ reported by Hsu and coworkers~\cite{Hsu2001}.

The phase behaviour of the system is understood in terms of a general electron-lattice coupling mechanism that we argue could apply to other members of the TMD family and, if strong enough, lead to stable orderings and/or miscibility in other binary compounds.
%
%
Comparison between 3D bulk and 2D convex hulls reveals interlayer coupling and system dimensionality, at the origin of sought-after exotic electronic behaviour, are negligible regarding phase stability of the binary alloys.
We argue that this finding should be valid for most 2D materials in which phase stability is governed by the similar in-plane electron-lattice effect, while more subtle behaviour could arise in presence of magnetic or Coulombic  interactions~\cite{Ramasubramaniam2013,Calandra2018}.
The possibility to limit phase diagrams to pure ML systems allows to work in a reduced combinatorial space and achieve a considerable speed-up of the protocol for future studies.
\appendix
\section{ECI value}\label{sec:ECI}
\begin{figure}[h]
  \includegraphics[width=0.85\textwidth]{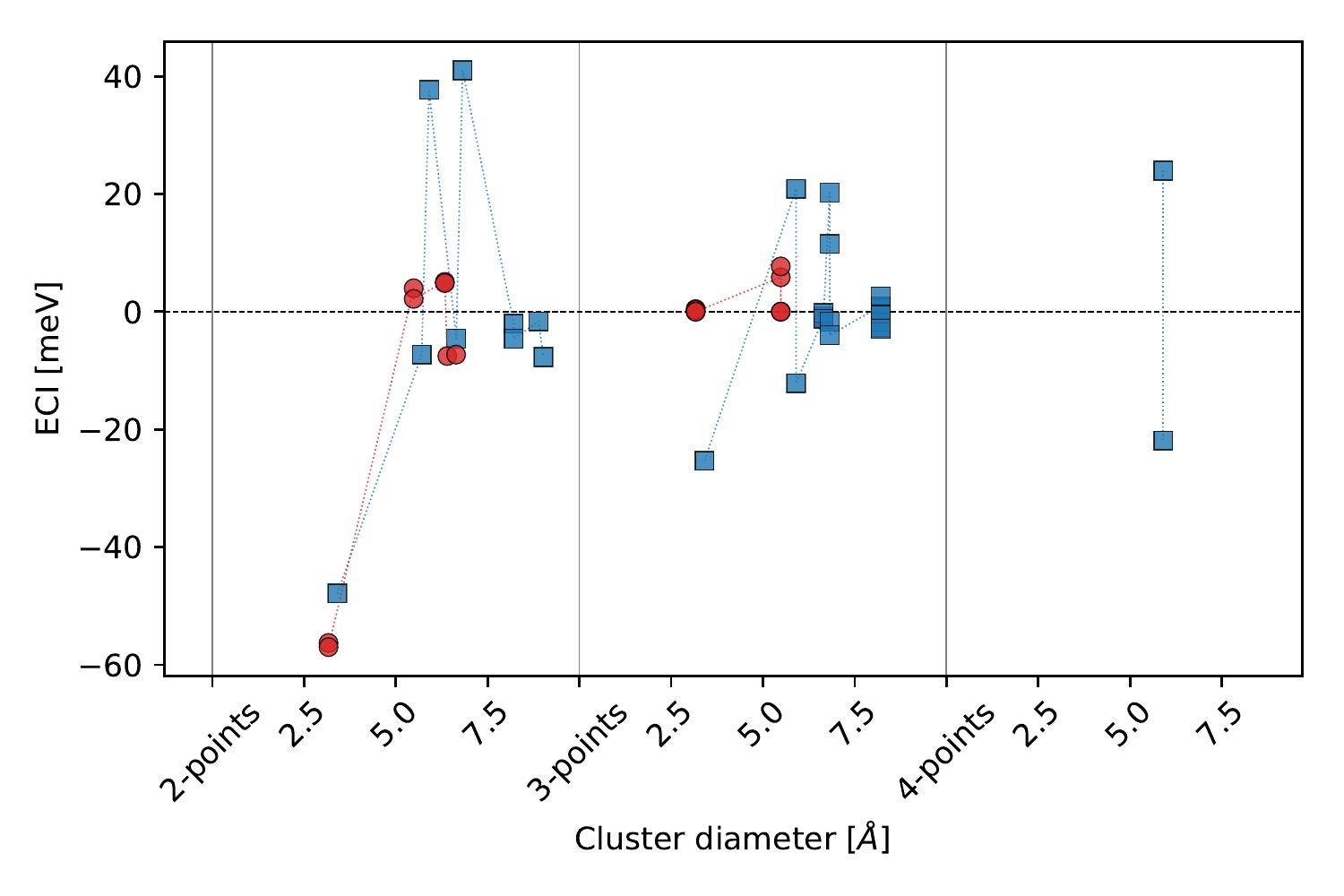}
  \caption{\label{fig:ECI}
  ECI in prismatic 2H host (red circles) and octahedral hosts (blue squares) as defined in Equation 1 in the main text.
  The dashed lines connecting the points are a guide to the eye.}
\end{figure}

\Cref{fig:ECI} shows the ECIs of the CE models used in this work.
ECI energies are plotted as a function of cluster diameter in groups of pairs, triplets, and quadruplets.
In both cases, the shortest range next-neighbour pair interaction has a negative ECI, favouring the same species neighbouring pairs. 
This interaction seems to be the leading term in the trigonal-prismatic host, as longer-range terms are significantly smaller in magnitude.
Conversely, in the octahedral host, longer-range pair interactions and higher-order clusters contribute similarly to the energy expansion Equation 1 in the main text. 

\section{Distortion of Host Lattice}\label{sec:host_dist}
\begin{figure}[h]
\centering
  \includegraphics[width=0.85\textwidth]{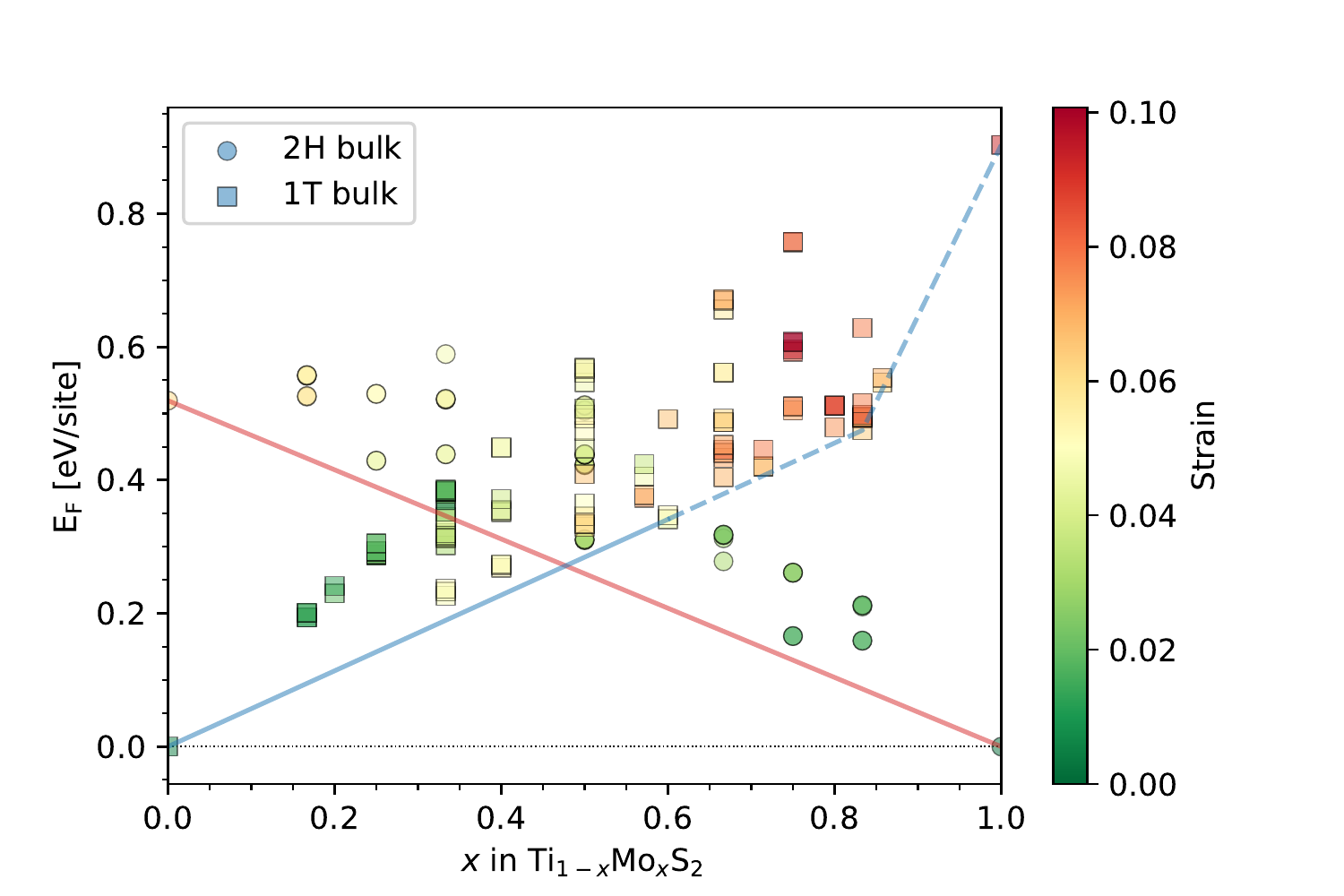}
  \caption{\label{fig:hull_strain}
  Distortion of all computed structure in 2H (circles) and 1T (squares) bulk system.
  The $y$ axis reports the formation energy per site versus the concentration $x$, like in Figure 2 in the main text. 
  The colour of each point shows the distortion value, increasing from no distortion (green) to the maximum distortion observed (red).
}
\end{figure}

\Cref{fig:hull_strain} reports the distortion from the native geometry occurring upon relaxation as a function of concentration $x$.
The host distortion is defined as the strain needed to transform the original cell into the relaxed one, apart from isotropic scaling and rotations.
Quantitatively, the distortion $\Delta$ is evaluated as the sum of the squared elements of the strain tensor~\cite{VandeWalle2002}:
$$
    \Delta = \sqrt{\sum_{i,j}\epsilon_{ij}^2}.
$$
%
%
Distortions $\Delta \geq 0.1$ are usually considered too large for the CE formalism to be applied~\cite{VandeWalle2002}, as the mapping of the relaxed configuration to the perfect lattice breaks down.

\section*{Acknowledgements}
This project has received funding from the European Union Horizon2020 research and innovation programme under grant agreement No. 721642: SOLUTION.
The authors acknowledge the use of the IRIDIS High Performance Computing Facility, and associated support services at the University of Southampton, in the completion of this work.
The authors are grateful to the UK Materials and Molecular Modelling Hub for computational resources, which is partially funded by EPSRC (EP/P020194/1).
TP acknowledges the support from the project OPVVV Novel nanostructures for engineering applications No. CZ.02.1.01/0.0/0.0/16\_026/0008396 supported by EU/MSMT.

\subparagraph{Author Contributions}
AS carried out all calculations and data analysis.
AS and DK conceptualize the study in the first place.
TP and DK supervised the study.
All authors contributed to the writing of the manuscript. 
%

\subparagraph{Competing interests} The authors declare no competing interests.

\bibliographystyle{elsarticle-num}

\end{document}